\documentclass{ws-procs9x6}
\newcommand{\pbar}{\ensuremath{\overline{\rm{p}}}}
\def\ssnn{\ensuremath{\sqrt{s_{\rm{NN}}}} }
\def\Raa{\ensuremath{R_{\rm{AuAu}}} }
\def\Rs{\ensuremath{R_{\rm{s}}} }
\def\Tc{\ensuremath{T_{\rm{c}}} }
\def\Rout{\ensuremath{R_{\rm{out}}} }
\def\Rside{\ensuremath{R_{\rm{side}}} }
\def\pt{\ensuremath{p_{\rm{t}}} }
\def\betat{\ensuremath{\beta_{\rm{t}}} }
\def\Ncoll{\ensuremath{N_{\rm{coll}}} }
\def\Journal#1#2#3#4{#1 {\bf #2}, #3 (#4)}

\def\JPG{J.~Phys.~G}

\def\NIMA{Nucl. Instrum. Methods~A}

\def\NPB{Nucl. Phys.~B}

\def\PRC{Phys. Rev.~C}

\def\APPB{Acta Phys. Pol.~B}
\begin{document}

\title{Systematics of Identified Hadron Spectra \\ at PHENIX}

\author{M. Csan\'ad\footnote{\textsc{This work was partly supported by the OTKA
T038406 grant}}\;\, for the PHENIX collaboration}

\maketitle

\abstracts{Mid-rapidity transverse momentum distributions for
$\pi^\pm$, $K^\pm$, p and $\pbar$ are measured by the PHENIX
experiment at RHIC in Au+Au, d+Au and p+p collisions at
\ssnn=200GeV up to ~2--4GeV. Also particle ratios of
$\pi^{-}/\pi^{+}$, $K^{-}/K^{+}$, $\pbar/p$, $p/\pi$ and
$\pbar/\pi$ are measured, as well as the nuclear modification
factor, all as a function of \pt and in every of the above
collision systems. Finally, the measured p+p and Au+Au spectra are
compared to the Buda-Lund hydro model.}

\section{Introduction} The motivation for ultra-relativistic
heavy-ion experiments at the Relativistic Heavy Ion Collider
(RHIC) at Brookhaven National Laboratory is the study of nuclear
matter at extremely high temperature and energy density with the
hope of creating and detecting deconfined matter consisting of
quarks and gluons -- the quark gluon plasma (QGP).

The PHENIX experiment at RHIC published spectra measurements in
Au+Au collisions\cite{PPG026} and in p+p and d+Au
collisions\cite{Felix-Brecken}, all at \ssnn=200GeV. In this
paper, we present a comparision of the \ssnn=200GeV Au+Au, d+Au
and p+p results.

\section{Measurements} The PHENIX experiment\cite{PHENIX_overview}
has a unique hadron identification capability in a broad momentum
range. Pions and kaons are identified up to 3~GeV/$c$ and
2~GeV/$c$ in $\pt$, respectively, and protons and anti-protons can
be identified up to 4.5~GeV/$c$ by using a high resolution
time-of-flight detector\cite{PHENIX_PID}.

We compare here identified particle production in d+Au, p+p and
peripheral and central Au+Au collisions.

First, we calculate the nuclear modification factor by taking the
Au+Au and d+Au spectra, scaling them down by the number of binary
nucleon-nucleon collisions (denoted as \Ncoll and taken from
Glauber calculations\cite{glauber} for each centrality class
separately) and dividing them by spectra measured in p+p
collisions (where \Ncoll is obviously one). Now, if a Au+Au or a
d+Au collision would be nothing but a combination of a lot of
nucleon-nucleon collisions, this ratio would be one. In contrary,
if there are effects due to the high $\Ncoll$, eg. some type of
medium is produced, this ratio can deviate from one. Furthermore,
from hydro, one would expect scaling with the number of
participants rather than the number of collisions.

\begin{figure}[ht]
\centerline{\epsfxsize=4.5in\epsfbox{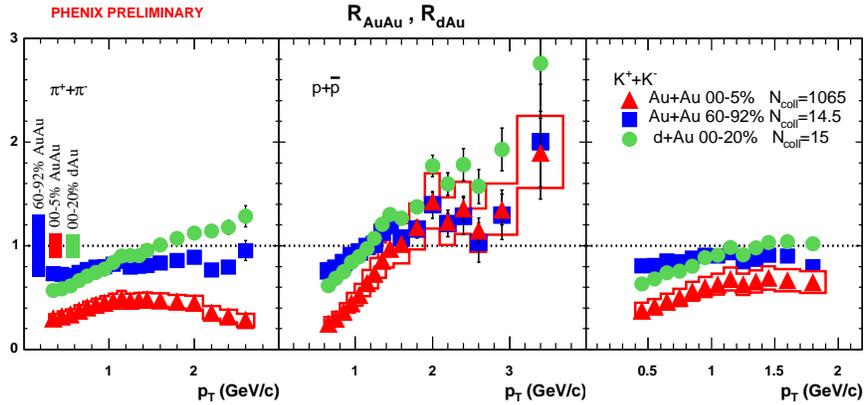}} \caption{Nuclear
modification factors for different particles and collision
species. Solid bars on the left indicate $\pt$-independent
normalization uncertainties, error bars indicate statistical
errors. Empty boxes indicate systematical errors.\label{f:nmf}}
\end{figure}

Fig.~\ref{f:nmf} compares the nuclear modification factors for
pions, kaons and (anti)protons in Au+Au and d+Au collisions. Pions
show a much lower \Raa at high \pt in central than in peripheral
Au+Au collisions, as expected from the large energy loss suffered
by the quarks in central collisions. The nuclear modification
factor is slightly larger in d+Au than in peripheral Au+Au,
despite the comparable number of binary collisions, but we can not
draw a definite conclusion due to the large systematic error of
the peripheral $\Raa$. The proton and antiproton nuclear
modification factors show a quite different trend, however.

\begin{figure}[ht]
\centerline{\epsfxsize=4.5in\epsfbox{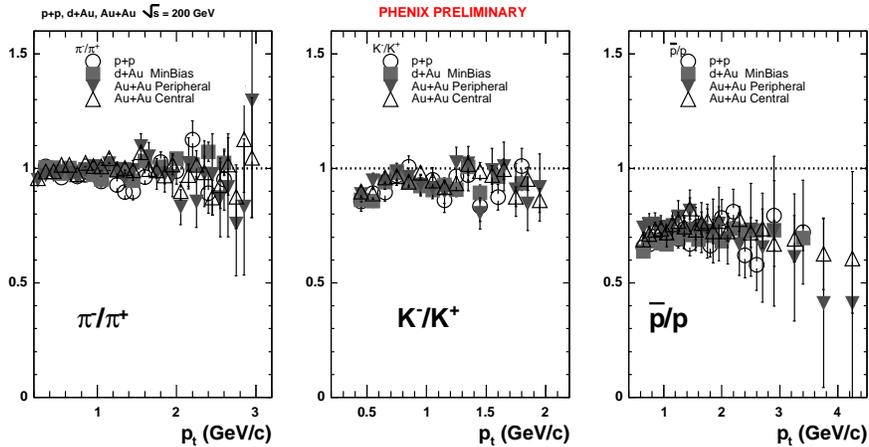}}
\caption{Proton to pion ratios for different collision
species\label{f:antipart}}
\end{figure}

Another important thing is to compare particle over antiparticle
ratios. What we see is that all collision species result in a
particle over antiparticle ratio of one, independently of
transverse momentum or centrality, except in the case of \pbar/p,
where the ratio is a bit smaller than one but still independent of
collision type and \pt as it is seen on Fig.~\ref{f:antipart}. We
can interpret the flat and collision species independent ratios of
one as a sign of thermalization, and the difference between
protons and antiprotons tells us that there is a small but nonzero
net barion density or bariochemical potential. The lack of
centrality dependence indicates that the (hadro-chemical)
freeze-out parameters are independent of the centrality of the
collisions\cite{PPG026,Rafelski}.

Last, but not least, we also calculate proton to pion ratios,
which will give us insight into collective dynamics effects, and
also tell us something about a possible barion yield enhancement
mechanism.

\begin{figure}[ht]
\centerline{\epsfxsize=2.25in\epsfbox{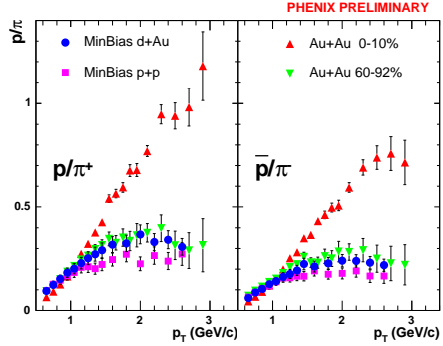}} \caption{The
ratio of protons to $\pi^+$ and antiprotons to $\pi^-$ in minimum
bias p+p and d+Au compared to peripheral and Au+Au collisions.
Statistical error bars are shown.\label{f:ptopi}}
\end{figure}

The p/$\pi$ ratio in d+Au is very similar to that in peripheral
Au+Au collisions, and lies slightly above the p+p ratio. The
p/$\pi$ ratio in central Au+Au collisions is, however, much
larger. All are plotted on Fig.~\ref{f:ptopi}. The difference
between the ratio in central and peripheral Au+Au clearly
indicates that baryon yield enhancement is not simply an effect of
sampling a large nucleus in the initial state, but it requires the
presence of a substantial volume of nuclear medium with high
energy density and pressure that generate a strong radial flow.

Now let us pick a model to compare it to our results. We choose
the Buda-Lund hydro model\cite{csorgo-blorig} which is successful
in describing the BRAHMS, PHENIX, PHOBOS and STAR data on
identified single particle spectra and the transverse mass
dependent Bose-Einstein or HBT radii (and so the
$\Rout/\Rside\approx 1$ behavior) as well as the pseudorapidity
distribution of charged particles in Au+Au collisions both at
\ssnn = 130 GeV\cite{ster-ismd03} and at \ssnn = 200
GeV\cite{csanad-qm04}.
\begin{figure}[ht]
\centerline{\epsfxsize=2.15in\epsfbox{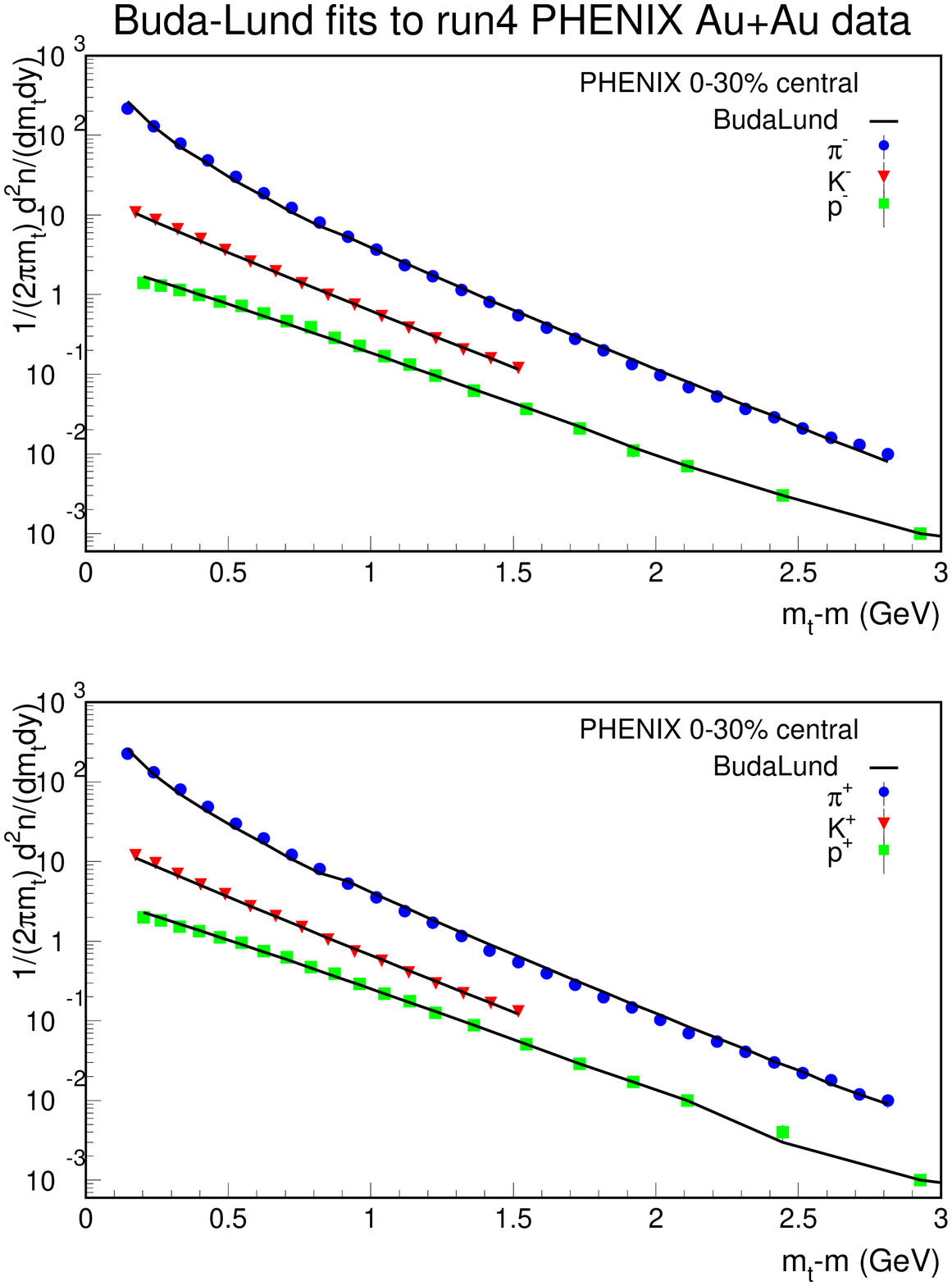}\epsfxsize=2.15in\epsfbox{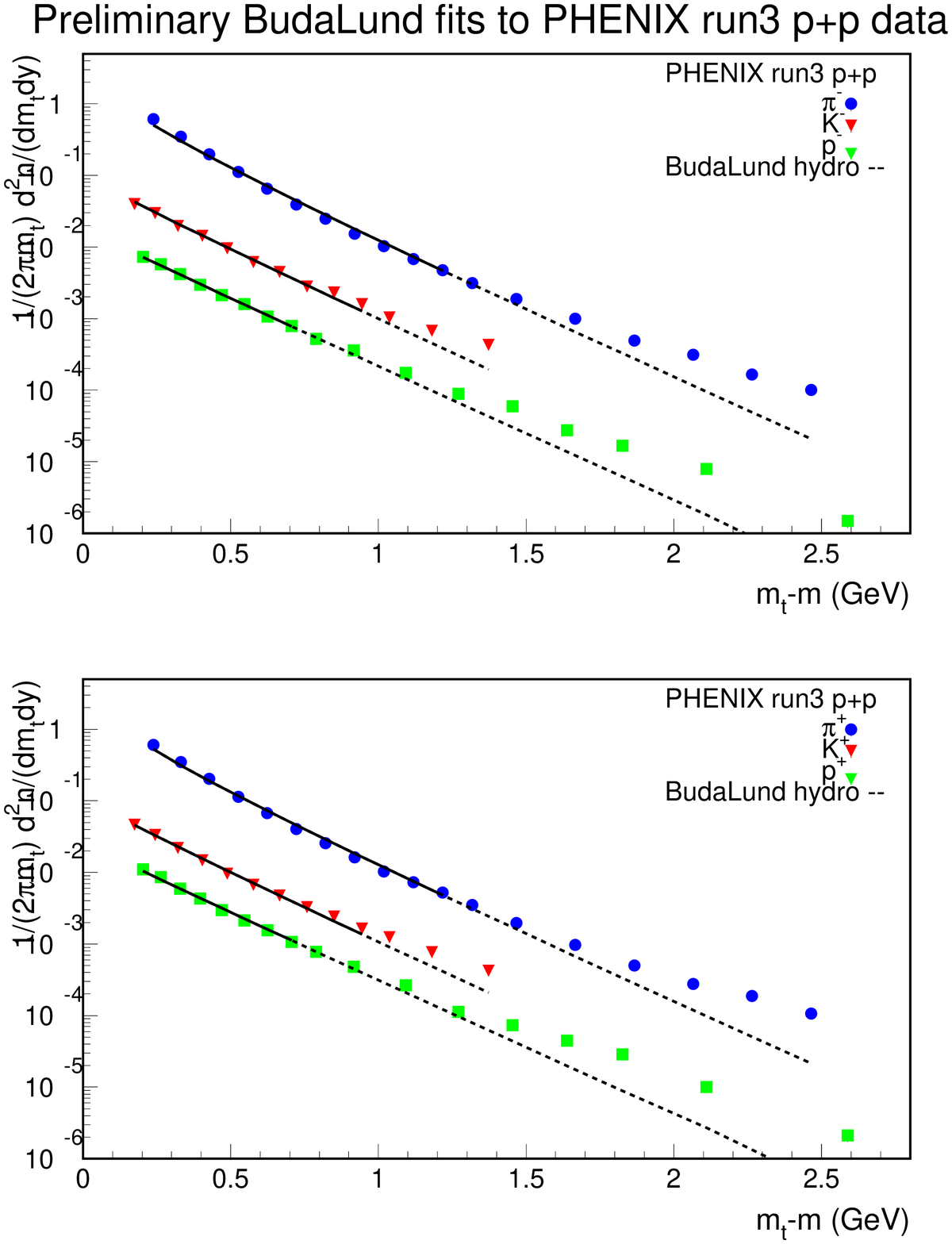}}
\caption{This figure shows a Buda-Lund fit to 200GeV Au+Au and p+p
data, latter is preliminary. The dashed line in the left panel is
a not fitted region, where jets start to dominate particle
production.\label{f:BL200}}
\end{figure}

In Au+Au collisions, we see a clear evidence for a 3-dimensional
Hubble-expansion, as the longitudinal and transverse flow is the
same\cite{csanad-qm04}.

The result of the Buda-Lund fits to p+p data at \ssnn = 200GeV
indicate that there is no radial flow ($\betat=0.0\pm0.19$) in p+p
collisions and the flow has a one-dimensional (Bjorken) flow
profile. This also means, that the spectra slopes correspond to
the temperature of the system.

Furthermore the temperature distribution $T(x)$ in the model is
characterized with a central temperature $T_0$ and a radius $\Rs$
where the temperature drops to the half of the central one. $T_0$
is found to be greater than the critical value calculated from
lattice QCD\cite{fodor-lattice}: $\Tc=164\pm 3$MeV, and
$T_0=196\pm 13$MeV in Au+Au\cite{csanad-qm04} and $T_0=239\pm
21$MeV in p+p. The Buda-Lund fits thus indicate quark
deconfinement at RHIC. \Rs is found to be finite and this is an
indication for temperature inhomogeneity and for the presence of a
cross-over instead of a phase transition, where no hadrons could
emerge from a (superheated) region with $T>\Tc$. In fact, the
temperature inhomogeneity also explains why $\Rout\approx
\Rside$\cite{mate-warsaw03}.

\section{Summary} The nuclear modification factor of pions, kaons and
also protons observed in d+Au is similar to the one observed in
peripheral Au+Au collisions. Pions are suppressed in central
Au+Au, and do not scale with $\Ncoll$. \Raa for protons and
antiprotons confirms previous observations that the production of
high \pt baryons in Au+Au scales with the number of binary
nucleon-nucleon collisions. Particle over antiparticle ratios are
near to one for pions and kaons, and slightly below one for
protons, independently of \pt and collision type. This can be
interpreted as a sign of thermalization. The proton to pion ratio
in p+p and d+Au is similar to that in peripheral Au+Au. In central
Au+Au we see a barion yield enhancement, which can be caused by
the presence of a hot and dense nuclear matter with a strong
radial flow. Buda-Lund fits to \pt spectra show us an evidence for
a 3d Hubble flow in Au+Au and a 1d Bjorken flow in p+p collisions,
and an indication of deconfinement temperatures reached in both
reactions.

\end{document}